\documentclass[aps,prl,twocolumn,superscriptaddress,showpacs]{revtex4}

\usepackage{graphicx}
\usepackage{dcolumn}
\usepackage{bm}

\begin{document}

\newlength{\ogltemp}

\def\met{\settowidth{\ogltemp}{\big\slash}%
\hspace{0.5mm}\big\slash\hspace{-1.4\ogltemp}{E_T}%
}
\def\metvet{\settowidth{\ogltemp}{\big\slash}%
\hspace{0.5mm}\big\slash\hspace{-1.4\ogltemp}{\vec{E}_T}%
}
\def\gev{\,\,\mathrm{GeV}\,}
\def\tev{\,\,\mathrm{TeV}\,}
\def\mev{\,\,\mathrm{MeV}\,}
\def\gevc{\,\,\mathrm{GeV}/c\,}
\def\gevm{\,\,\mathrm{GeV}/c^2\,}
\def\D0   {${\rm D}{\rm O\!\!\!\!\hspace{0.5pt}\protect\raisebox{0.2ex}{/}}$ }

\title{Search for Gluino-Mediated Bottom Squark Production in $p\bar{p}$ Collisions at $\sqrt{s}=1.96$~TeV}

\affiliation{Institute of Physics, Academia Sinica, Taipei, Taiwan 11529, Republic of China} 
\affiliation{Argonne National Laboratory, Argonne, Illinois 60439} 
\affiliation{University of Athens, 157 71 Athens, Greece} 
\affiliation{Institut de Fisica d'Altes Energies, Universitat Autonoma de Barcelona, E-08193, Bellaterra (Barcelona), Spain} 
\affiliation{Baylor University, Waco, Texas  76798} 
\affiliation{Istituto Nazionale di Fisica Nucleare Bologna, $^w$University of Bologna, I-40127 Bologna, Italy} 
\affiliation{Brandeis University, Waltham, Massachusetts 02254} 
\affiliation{University of California, Davis, Davis, California  95616} 
\affiliation{University of California, Los Angeles, Los Angeles, California  90024} 
\affiliation{University of California, San Diego, La Jolla, California  92093} 
\affiliation{University of California, Santa Barbara, Santa Barbara, California 93106} 
\affiliation{Instituto de Fisica de Cantabria, CSIC-University of Cantabria, 39005 Santander, Spain} 
\affiliation{Carnegie Mellon University, Pittsburgh, PA  15213} 
\affiliation{Enrico Fermi Institute, University of Chicago, Chicago, Illinois 60637} 
\affiliation{Comenius University, 842 48 Bratislava, Slovakia; Institute of Experimental Physics, 040 01 Kosice, Slovakia} 
\affiliation{Joint Institute for Nuclear Research, RU-141980 Dubna, Russia} 
\affiliation{Duke University, Durham, North Carolina  27708} 
\affiliation{Fermi National Accelerator Laboratory, Batavia, Illinois 60510} 
\affiliation{University of Florida, Gainesville, Florida  32611} 
\affiliation{Laboratori Nazionali di Frascati, Istituto Nazionale di Fisica Nucleare, I-00044 Frascati, Italy} 
\affiliation{University of Geneva, CH-1211 Geneva 4, Switzerland} 
\affiliation{Glasgow University, Glasgow G12 8QQ, United Kingdom} 
\affiliation{Harvard University, Cambridge, Massachusetts 02138} 
\affiliation{Division of High Energy Physics, Department of Physics, University of Helsinki and Helsinki Institute of Physics, FIN-00014, Helsinki, Finland} 
\affiliation{University of Illinois, Urbana, Illinois 61801} 
\affiliation{The Johns Hopkins University, Baltimore, Maryland 21218} 
\affiliation{Institut f\"{u}r Experimentelle Kernphysik, Universit\"{a}t Karlsruhe, 76128 Karlsruhe, Germany} 
\affiliation{Center for High Energy Physics: Kyungpook National University, Daegu 702-701, Korea; Seoul National University, Seoul 151-742, Korea; Sungkyunkwan University, Suwon 440-746, Korea; Korea Institute of Science and Technology Information, Daejeon, 305-806, Korea; Chonnam National University, Gwangju, 500-757, Korea} 
\affiliation{Ernest Orlando Lawrence Berkeley National Laboratory, Berkeley, California 94720} 
\affiliation{University of Liverpool, Liverpool L69 7ZE, United Kingdom} 
\affiliation{University College London, London WC1E 6BT, United Kingdom} 
\affiliation{Centro de Investigaciones Energeticas Medioambientales y Tecnologicas, E-28040 Madrid, Spain} 
\affiliation{Massachusetts Institute of Technology, Cambridge, Massachusetts  02139} 
\affiliation{Institute of Particle Physics: McGill University, Montr\'{e}al, Canada H3A~2T8; and University of Toronto, Toronto, Canada M5S~1A7} 
\affiliation{University of Michigan, Ann Arbor, Michigan 48109} 
\affiliation{Michigan State University, East Lansing, Michigan  48824}
\affiliation{Institution for Theoretical and Experimental Physics, ITEP, Moscow 117259, Russia} 
\affiliation{University of New Mexico, Albuquerque, New Mexico 87131} 
\affiliation{Northwestern University, Evanston, Illinois  60208} 
\affiliation{The Ohio State University, Columbus, Ohio  43210} 
\affiliation{Okayama University, Okayama 700-8530, Japan} 
\affiliation{Osaka City University, Osaka 588, Japan} 
\affiliation{University of Oxford, Oxford OX1 3RH, United Kingdom} 
\affiliation{Istituto Nazionale di Fisica Nucleare, Sezione di Padova-Trento, $^x$University of Padova, I-35131 Padova, Italy} 
\affiliation{LPNHE, Universite Pierre et Marie Curie/IN2P3-CNRS, UMR7585, Paris, F-75252 France} 
\affiliation{University of Pennsylvania, Philadelphia, Pennsylvania 19104}
\affiliation{Istituto Nazionale di Fisica Nucleare Pisa, $^y$University of Pisa, $^z$University of Siena and $^{aa}$Scuola Normale Superiore, I-56127 Pisa, Italy} 
\affiliation{University of Pittsburgh, Pittsburgh, Pennsylvania 15260} 
\affiliation{Purdue University, West Lafayette, Indiana 47907} 
\affiliation{University of Rochester, Rochester, New York 14627} 
\affiliation{The Rockefeller University, New York, New York 10021} 
\affiliation{Istituto Nazionale di Fisica Nucleare, Sezione di Roma 1, $^{bb}$Sapienza Universit\`{a} di Roma, I-00185 Roma, Italy} 

\affiliation{Rutgers University, Piscataway, New Jersey 08855} 
\affiliation{Texas A\&M University, College Station, Texas 77843} 
\affiliation{Istituto Nazionale di Fisica Nucleare Trieste/Udine, $^{cc}$University of Trieste/Udine, Italy} 
\affiliation{University of Tsukuba, Tsukuba, Ibaraki 305, Japan} 
\affiliation{Tufts University, Medford, Massachusetts 02155} 
\affiliation{Waseda University, Tokyo 169, Japan} 
\affiliation{Wayne State University, Detroit, Michigan  48201} 
\affiliation{University of Wisconsin, Madison, Wisconsin 53706} 
\affiliation{Yale University, New Haven, Connecticut 06520} 
\author{T.~Aaltonen}
\affiliation{Division of High Energy Physics, Department of Physics, University of Helsinki and Helsinki Institute of Physics, FIN-00014, Helsinki, Finland}
\author{J.~Adelman}
\affiliation{Enrico Fermi Institute, University of Chicago, Chicago, Illinois 60637}
\author{T.~Akimoto}
\affiliation{University of Tsukuba, Tsukuba, Ibaraki 305, Japan}
\author{M.G.~Albrow}
\affiliation{Fermi National Accelerator Laboratory, Batavia, Illinois 60510}
\author{B.~\'{A}lvarez~Gonz\'{a}lez}
\affiliation{Instituto de Fisica de Cantabria, CSIC-University of Cantabria, 39005 Santander, Spain}
\author{S.~Amerio$^x$}
\affiliation{Istituto Nazionale di Fisica Nucleare, Sezione di Padova-Trento, $^x$University of Padova, I-35131 Padova, Italy} 

\author{D.~Amidei}
\affiliation{University of Michigan, Ann Arbor, Michigan 48109}
\author{A.~Anastassov}
\affiliation{Northwestern University, Evanston, Illinois  60208}
\author{A.~Annovi}
\affiliation{Laboratori Nazionali di Frascati, Istituto Nazionale di Fisica Nucleare, I-00044 Frascati, Italy}
\author{J.~Antos}
\affiliation{Comenius University, 842 48 Bratislava, Slovakia; Institute of Experimental Physics, 040 01 Kosice, Slovakia}
% ...

\collaboration{CDF Collaboration\footnote{With visitors from $^a$University of Massachusetts Amherst, Amherst, Massachusetts 01003,
$^b$Universiteit Antwerpen, B-2610 Antwerp, Belgium, 
$^c$University of Bristol, Bristol BS8 1TL, United Kingdom,
$^d$Chinese Academy of Sciences, Beijing 100864, China, 
$^e$Istituto Nazionale di Fisica Nucleare, Sezione di Cagliari, 09042 Monserrato (Cagliari), Italy,
$^f$University of California Irvine, Irvine, CA  92697, 
$^g$University of California Santa Cruz, Santa Cruz, CA  95064, 
$^h$Cornell University, Ithaca, NY  14853, 
$^i$University of Cyprus, Nicosia CY-1678, Cyprus, 
$^j$University College Dublin, Dublin 4, Ireland,
$^k$Royal Society of Edinburgh/Scottish Executive Support Research Fellow,
$^l$University of Edinburgh, Edinburgh EH9 3JZ, United Kingdom, 
$^m$Universidad Iberoamericana, Mexico D.F., Mexico,
$^n$Queen Mary, University of London, London, E1 4NS, England,
$^o$University of Manchester, Manchester M13 9PL, England, 
$^p$Nagasaki Institute of Applied Science, Nagasaki, Japan, 
$^q$University of Notre Dame, Notre Dame, IN 46556,
$^r$University de Oviedo, E-33007 Oviedo, Spain, 
$^s$Simon Fraser University, Vancouver, British Columbia, Canada V6B 5K3,
$^t$Texas Tech University, Lubbock, TX  79409, 
$^u$IFIC(CSIC-Universitat de Valencia), 46071 Valencia, Spain,
$^v$University of Virginia, Charlottesville, VA  22904,
$^{dd}$On leave from J.~Stefan Institute, Ljubljana, Slovenia, 
}}
\noaffiliation

\vfill\eject

\date{\today}

\begin{abstract}
We report on a search for the supersymmetric partner of the bottom quark produced from gluino 
decays in data from 2.5~fb$^{-1}$ of integrated luminosity collected by the Collider Detector
at Fermilab at $\sqrt{s}=1.96$~TeV. Candidate events are selected requiring two or more 
jets and large missing transverse energy. At least two of the jets are required to be 
tagged as originating from a $b$ quark to enhance the sensitivity. The results are in 
good agreement with the prediction of the standard model processes, giving no evidence 
for gluino decay to sbottom quarks. This result constrains the gluino-pair-production cross 
section to be less than 40~fb at 95~\% credibility level for a gluino mass of $350 \gevm$.
\end{abstract}


\maketitle

The standard model (SM) of elementary particles and fundamental 
interactions, however successful, is incomplete, since it
does not explain the origin of electroweak symmetry breaking or
the gauge hierarchy problem~\cite{Gildener:1976}.
A proposed extension of the SM, supersymmetry
(SUSY)~\cite{Martin:1997ns}, solves these problems by introducing 
a symmetry that relates particles of different spin. 
R-parity~\cite{Martin:1997ns} conserving SUSY models also
provide a prime candidate for the dark matter in the
cosmos~\cite{Goldberg:1983}, namely the stable lightest
supersymmetric particle (LSP). In these models, the
left-handed and right-handed quarks have scalar partners denoted
$\tilde{q}_L$ and $\tilde{q}_R$ which can mix to form scalar
quarks (squarks) with mass eigenstates $\tilde{q}_{1,2}$. Several
models~\cite{Bartl:1994bu} predict that this mixing can be
substantial for the scalar bottom (sbottom), yielding a sbottom
mass eigenstate ($\tilde{b}$) significantly lighter than other
squarks. In proton-antiproton ($p \bar{p}$) collisions at the Tevatron's
 center-of-mass energy of $\sqrt{s}=1.96\tev$, the gluino 
($\tilde{g}$, the spin-1/2 superpartner of the gluon) pair-production cross 
section is almost an order of magnitude larger than that of a sbottom of similar
mass~\cite{Beenakker:1996ch}. Therefore, if sufficiently
light, sbottom quarks could be copiously produced through the
$\tilde{g} \to \tilde{b}b$ decays since the gluino preferentially 
decays into a squark-quark pair~\cite{Martin:1997ns}.
A sbottom in the mass range accessible at the
Tevatron is expected to decay predominantly into a bottom quark
and the lightest neutralino ($\tilde{\chi}^{0}$), which is often assumed to be the LSP.
Previous searches for direct
sbottom~\cite{stoplbjetrun1paper,dzerosbottom2006} or gluino
production~\cite{squarkgluinorun1paper,carsten} at the Tevatron
placed lower limits on the masses of these particles. 

In this Letter, we report the search for $\tilde{g} \to \tilde{b}b$ 
decays in $p\bar{p}$ collision data from $2.5$~fb$^{-1}$ of integrated 
luminosity collected between March~2003 and April~2008 by the upgraded 
Collider Detector at Fermilab (CDF II) at the Tevatron.
Assuming \mbox{R-parity} conservation, $\tilde{g}$'s
are produced in pairs. We consider a scenario
where the branching fractions $\tilde{g} \rightarrow b\tilde{b}$
and $\tilde{b} \rightarrow b\tilde{\chi}^{0}$ are $100\%$.
If these conditions are satisfied the analysis is largely independent 
of the $\tilde{\chi}^{0}$ mass, as long as the $m(\tilde{b}) - m(\tilde{\chi}^{0})$
is larger than $25 \gevm$, due to b-jet energy cut (to be discussed later). For our calculations
we assume a $\tilde{\chi}^{0}$ mass of $60\gevm$, which is above the
limits from LEP~\cite{neutralino_mass}. 
The final state contains four $b$ jets from
the hadronization of the $b$ quarks and an imbalance in momentum in the
transverse plane to the beam (``missing transverse energy" or
$\met$~\cite{coordinates}) from the two undetected LSPs. 

CDF II is a multipurpose detector, described in detail
elsewhere~\cite{CDFII}.
The charged-particle tracking system consists of silicon microstrip detectors
and a cylindrical open-cell drift chamber in a
$1.4$~T solenoidal magnetic field coaxial with the beam line.
The silicon detectors provide coverage in the 
pseudorapidity~\cite{coordinates} range $|\eta| \le 2$ and 
are used to identify events with long-lived particle decays. 
The drift chamber surrounds the silicon detectors 
and has maximum efficiency up to $|\eta| = 1$.
Segmented sampling calorimeters, arranged in a projective tower 
geometry, surround the tracking system, and measure the energy 
of interacting particles for $|\eta| \le 3.6$. Muons
are identified by drift chambers, which extend to $|\eta| = 1.5$, 
and are located outside the calorimeter volume.
Jets are reconstructed from the energy depositions in the
calorimeter cells using an iterative cone jet-clustering
algorithm~\cite{Arnison:1983rn}, with a cone size of
radius \mbox{$R = \sqrt{(\Delta\phi)^2+(\Delta \eta)^2} = 0.4$}~\cite{coordinates}.
Energy corrections~\cite{JES} are applied to account for effects that 
distort the measured jet energy, such as non-linear calorimeter
response, underlying event, and the position of the primary
vertex.

Candidate events used for this search are selected by an
on-line event selection system, a (trigger) requiring $\met\ge45\gev$.
Further selections remove accelerator-produced and
detector-related backgrounds as well as cosmic-ray events.
After off-line event reconstruction, the events are required to have $\met \ge 70$~GeV, and at least two
jets with $|\eta| \le 2.4$ and $E_T \ge 25$~GeV. The highest-$E_T$ jet is required 
to have $E_T \ge 35$~GeV and at least one of the selected jets is required to 
have $|\eta| \le 0.9$.
The $B$ hadrons in jets coming from b quark fragmentation have an average flight path of about 
500 microns, yielding secondary vertices relative to the interaction point (primary vertex).
We require the events to have at least two jets identified as $b$ jets 
by the CDF secondary-vertex $b$-tagging
algorithm~\cite{Acosta:2004hw}. The double $b$-tagging requirement 
effectively enhances the sensitivity. 

Dominant SM backgrounds are top-quark pair-production and single top-quark production, 
electroweak boson and diboson production, heavy-flavor (HF) multijet production, and light-flavor jets 
falsely tagged as $b$ jets (mistags). The latter two background contributions are 
estimated from data. The \textsc{pythia} event generator~\cite{Sjostrand:2000wi} is used 
to estimate the remaining backgrounds. For the event generation the CTEQ5L~\cite{CTEQ5} 
parton distribution functions were used. Events are passed through the 
\textsc{geant3}-based~\cite{GEANT3} CDF II detector simulation~\cite{simulation} 
and weighted by the probability that they would pass the trigger as determined in 
independent data samples. The single top-quark and diboson event yields are normalized 
to the theoretical cross sections~\cite{xsec_singtop,xsec_diboson1,xsec_diboson2}. 
The event yields for the electroweak boson samples 
are normalized to the leading order cross section provided by \textsc{pythia}, scaled by a factor 
of 1.4 to account for higher order corrections. We use the top-quark pair production cross section 
of $\sigma_{t\bar{t}} = 7.3 \pm 0.8$~pb~\cite{top_xsec}. 
Mistags are estimated from inclusive jet-sample data by computing 
a mistag rate~\cite{Acosta:2004hw} which is parametrized by the jet $E_{T}$, $|\eta|$,
secondary-vertex track-multiplicity, the number of primary vertices in the event,
primary vertex $z$-position, and the scalar sum of $E_{T}$ of all jets in the event.
To estimate the HF multijet background from data, we have developed a multijet tag rate 
estimator (MUTARE). The estimator is based on a tag-rate matrix applied to each 
jet in an event following a parametrization of $E_{T}$, $|\eta|$ and the scalar sum of $E_{T}$ of all 
jets in the event. Each element of the matrix is computed in a reference sample as the ratio between the 
number of $b$-tagged jets minus the number of mistags over the number of ``taggable'' jets, 
where ``taggable'' jets are defined as jets with tracks passing the CDF secondary-vertex 
$b$-tagging algorithm requirements~\cite{Acosta:2004hw}.
The final prediction is obtained after subtracting the HF contribution coming 
from non-multijet production processes.   
The amount of non-multijet HF contribution is computed 
by applying the MUTARE matrix to each non-multijet background sample described above.

\begin{figure}

\includegraphics[width=7.5cm]{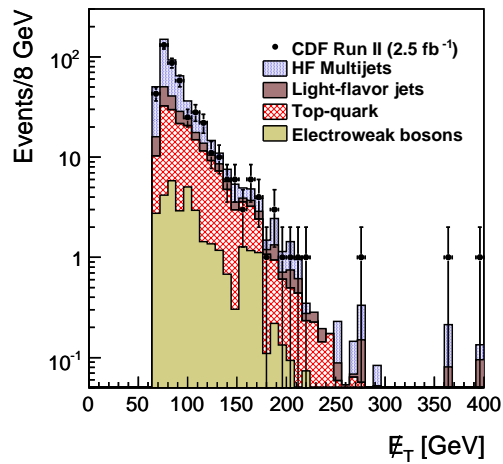}
\caption{ \label{met} Distribution of $\met$ in the pre-optimization region in which leading $E_{T}$ and 
  second-leading $E_{T}$ jets are not aligned with the $\met$, and isolated leptons 
  are vetoed. SM prediction (stacked histograms) and observed distribution (dots) 
  are shown, where HF multijets and light-flavor jets are predicted from data as an integrated estimation.}
\end{figure}

To avoid potential biases when searching for new physics, we test
the various background contributions in distinct
control regions that are defined {\it a priori}. 
The three control regions used to check the SM
prediction are denoted as multijet, lepton, and pre-optimization regions.
The multijet control region is defined to have the second leading $E_T$ jet ($\vec{j}_{2}$) aligned with the $\metvet$~\cite{coordinates},
where aligned means $\Delta \phi (\metvet,\vec{j}_{2}) \leq 0.4$~rad.
This HF multijet enriched region is used to build the MUTARE matrix to predict 
the HF multijet background in the other control and signal regions.  
The lepton control region is defined to have $\vec{j}_{2}$ not aligned with the 
$\metvet$ ($\Delta \phi (\metvet,\vec{j}_{2}) \geq 0.7$~rad) and at least one isolated lepton 
with $p_{T} \geq 10 \gevc$. This lepton region
is used to check the top quark and electroweak $W/Z$ boson backgrounds. The pre-optimization control region is 
defined to have the leading and second-leading $E_T$ jets not aligned with the $\metvet$ 
and to have no identified leptons. Predicted total numbers of events 
and distributions of kinematic variables such
as jet $E_T$, the track multiplicity, and the $\met$ have been
studied and found to be in agreement with
observations in the three control regions. As
an example, the $\met$ distributions for the pre-optimization region are shown 
in Fig.~\ref{met}. The background contributions to
the number of expected inclusive double $b$-tagged events and the
observed events in the control regions are summarized in Table~\ref{tab:table1}.

\begin{table}
  \caption{\label{tab:table1} Comparison of the total number of expected events with total
    uncertainties and observed double $b$-tagged events in the control regions.}
  \begin{tabular}{c c c c}
    \hline
  \hline
    Regions:                        & Multijet            &  Lepton        & Pre-optimization \\
    \hline
       Electroweak bosons           & $10 \pm 7$     & $21 \pm 14$    & $33 \pm 22$ \\
       Top-quark                    & $19  \pm 6$    & $111 \pm 34$   & $146\pm 45$  \\
       Light-flavor jets            & $225 \pm 49$   & $8  \pm 2$     & $57 \pm 12$ \\
       HF Multijets                 & $839 \pm 419$  & $25 \pm 12$    & $270 \pm 135$ \\
   \hline
     Total expected                 & $1093 \pm 422$ & $165 \pm 39$   & $506 \pm 144$ \\
   \hline
     Observed                       & 1069           & 159            & 451      \\
     \hline
     \hline
\end{tabular}
\end{table}

We optimize the sensitivity to sbottom production from gluino
decays by using two Neural Networks (NN) trained with the \textsc{tmva} package~\cite{TMVA}. 
One of them is optimized to remove the HF multijet background (multijet-NN) and the 
other to remove the top-quark pair background (top-NN). 
The training is based on jet variables using, jet $E_T$, $\Delta \phi (\metvet,\vec{j}_{i})$, 
$\met$, and the summed $E_T$ of all the jets in the event.
We choose two reference signal points based on values of 
$\Delta m \equiv m(\tilde{g}) - m(\tilde{b})$ and perform the same optimization 
procedure. We refer to large $\Delta m$ optimization 
with $m(\tilde{g}) = 335 \gevm$ and $m(\tilde{b}) = 260 \gevm$ and to small 
$\Delta m$ optimization with $m(\tilde{g}) = 335 \gevm$ and $m(\tilde{b}) = 315 \gevm$.
These two signal points represent two different kinematic regions. 
For the large $\Delta m$ optimization three or more jets are required before applying 
the NN procedure. For the small $\Delta m$ optimization two or more jets are required 
since the $E_{T}$ spectrum of the $b$ jets is much softer. The signal predictions are 
obtained by computing the acceptance using the \textsc{pythia} event generator 
normalized to the NLO production cross section determined with \textsc{prospino} 
event generator~\cite{Beenakker:1996ch} and the CTEQ6M~\cite{CTEQ6_1,CTEQ6_2} parton 
distribution functions. The uncertainty of the NLO production cross section varies 
from 20\% ($m(\tilde{g}) \simeq 200 \gevm$) to 30\% ($m(\tilde{g}) \simeq 400 \gevm$). 

The systematic uncertainties on the signal and the background
predictions, taking into account correlated and uncorrelated uncertainties, are studied.
Correlated uncertainties, affecting both the background prediction
and signal acceptance, are dominated by the jet energy scale 
($16\%$ ($25\%$)~\cite{JES} for the large (small) $\Delta m$ optimization region), the different
$b$-tagging efficiency between data and simulation ($4.4\%$ ($4.9\%$)~\cite{Acosta:2004hw} for the 
large~(small) $\Delta m$ optimization region), and the luminosity~($6\%$)~\cite{CDFII}.
Uncorrelated systematic uncertainties on the
background predictions are dominated by the HF multijet $b$-tag rate ($50\%$), 
the mistag rate ($16\%$~\cite{Acosta:2004hw} for light-flavor multijets), 
the top-quark pair-production cross section ($11\%$), the single top-quark production cross section ($13\%$), 
and the diboson production cross section ($10\%$ for $WW$/$WZ$ and $20\%$ for $ZZ$). 
Due to the limited ability of \textsc{pythia} to simulate multijet environments, a $40\%$
uncertainty~\cite{prl_viktor} is assigned for the extracted yields of events with a $W$ or $Z$ boson and jets. 
Correlated and uncorrelated uncertainties are evaluated separately and combined in quadrature.

\begin{figure}
\includegraphics[width=7.5cm]{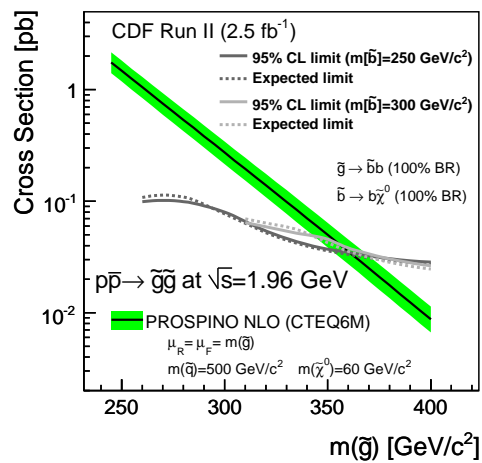}
\caption{Observed (solid lines) and expected (dashed lines) 95\% C.L. upper limit 
on the gluino cross section (solid line with band) as a function of the gluino mass for two assumed values 
of the sbottom mass. The shaded band denotes the uncertainty on the NLO $\tilde{g}\tilde{g}$ 
production due to the truncated higher-order terms and the parton distribution functions.}
\label{fig:limit_cs}
\end{figure}

The signal region is analyzed after the background predictions
are determined. We find 0.8 as an optimal value for the selection cut for both multijet-NN outputs and    
0.6 (0.8) for the top-NN outputs in the large (small) $\Delta m$ optimization within an 
interval of $-$1 to 1, where the background peaks at $-$1 and the signal peaks at 1. 
We observe 5 (2) events for the large (small) $\Delta m$ optimization region, 
where $4.7\pm 1.5$ ($2.4\pm 0.8$) are expected from background, as summarized 
in Table~\ref{tab:table2}. 
Since no significant deviation from the SM prediction is observed, the results are used
to calculate an exclusion limit for the cross section of the described gluino process.
We use a Bayesian method to determine the 95\% credibility level (C.L.) upper limit on the 
$\tilde{g}\tilde{g}$ cross section, assuming a uniform prior probability density. 
We treat the various correlated uncertainties as nuisance parameters, which we 
remove by marginalization, assuming a Gaussian prior distribution. 
The obtained limit is such that no more than 8.0~(5.4) 
events are observed in the large (small) $\Delta m$ signal region. 
Figure~\ref{fig:limit_cs} shows the expected and observed limits as a function of 
$m(\tilde{g})$ for two values of the $\tilde{b}$ quark mass. The expected limit 
is computed by assuming that the observed number of events matches 
the SM expectation in each signal region.

\begin{table}
 
\caption{\label{tab:table2} Number of expected and observed events
in the signal regions. Predictions for the signal
points are also shown. Correlated and uncorrelated uncertainties in the total
background and expected signal were treated separately in the analysis 
although they are combined here.}
\begin{tabular}{c c c}
  \hline
  \hline
  Optimizations:                    & Large $\Delta m$  & Small $\Delta m$ \\
  \hline
       Electroweak bosons           & $0.17 \pm 0.05$   & $0.5 \pm 0.3$\\ 
       Top-quark                    & $1.9 \pm 1.0$     & $0.6 \pm 0.4$\\
       Light-flavor jets            & $1.0 \pm 0.3$     & $0.6 \pm 0.1$\\
       HF Multijets                 & $1.6 \pm 0.8$     & $0.7 \pm 0.3$\\
   \hline
     Total expected SM              & $4.7 \pm 1.5$     & $2.4 \pm 0.8$\\
   \hline
     Observed                       & 5                 & 2   \\
     \hline
     \hline
     Optimized $\tilde{g}$ signal   & $14.9 \pm 5.0$    & $8.5 \pm 2.8$   \\
     \hline
     \hline
\end{tabular}
\end{table}

\begin{figure}
\includegraphics[width=7.5cm]{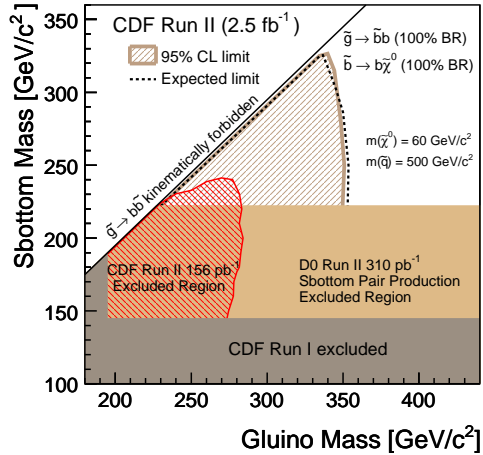}
\caption{Excluded region at 95\% C.L. in the $m(\tilde{g})$-$m(\tilde{b})$ plane 
  for a $m(\tilde{\chi}^{0}) = 60 \gevm$, $m(\tilde{q}) = 500 \gevm$. The result is compared to
  the previous results from CDF in Run I~\cite{squarkgluinorun1paper}, 
  and Run II~\cite{carsten} and direct sbottom production by D0~\cite{dzerosbottom2006}.}
\label{fig:limit_mass}
\end{figure}

The gluino production cross section limit is nearly independent of the sbottom mass
between 250 and 300~$\gevm$, and is around 40~fb for $m(\tilde{g})~=~350\gevm$. 
In addition, using the assumed model, a 95\%~C.L. limit is 
obtained in the parameter plane of the model.
Figure~\ref{fig:limit_mass} shows the excluded region in the gluino-sbottom mass plane, 
compared  with the results from previous 
analyses~\cite{dzerosbottom2006,stoplbjetrun1paper,squarkgluinorun1paper,carsten}.
The limit obtained with the present analysis improves the results of previous
searches using similar topology and also, under the assumptions discussed above, sets a more 
stringent limit on the sbottom and gluino production than dedicated sbottom searches.

In conclusion, we have searched for sbottom quarks from gluino
decays in $2.5$~fb$^{-1}$ of CDF Run II data. We observe 5 (2)
inclusive double $b$-tagged candidate events for the large (small) $\Delta m$ optimization region, which is in
agreement with SM background expectations of $4.7\pm 1.5$ ($2.4\pm 0.8$) events.
No evidence for sbottom quarks from gluino decays is observed, and 
we exclude a significant region in the gluino and sbottom
mass plane at $95\%$~C.L. 
For the assumed model, the limit is nearly independent of the sbottom mass and the cross section
limit is around 40~fb for $m(\tilde{g}) = 350\gevm$.

We thank the Fermilab staff and the technical staffs of the participating institutions 
for their vital contributions. This work was supported by the U.S. Department of Energy 
and National Science Foundation; the Italian Istituto Nazionale di Fisica Nucleare; the 
Ministry of Education, Culture, Sports, Science and Technology of Japan; the Natural 
Sciences and Engineering Research Council of Canada; the National Science Council of the 
Republic of China; the Swiss National Science Foundation; the A.P. Sloan Foundation; the 
Bundesministerium f\"ur Bildung und Forschung, Germany; the Korean Science and Engineering 
Foundation and the Korean Research Foundation; the Science and Technology Facilities 
Council and the Royal Society, UK; the Institut National de Physique Nucleaire et Physique 
des Particules/CNRS; the Russian Foundation for Basic Research; the Ministerio de Ciencia 
e Innovaci\'{o}n, and Programa Consolider-Ingenio 2010, Spain; the Slovak R\&D Agency; 
and the Academy of Finland. 

\end{document}